\documentclass[a4paper, 11pt]{article}
\RequirePackage[OT1]{fontenc}
\RequirePackage{natbib}
\usepackage{amsfonts,amssymb,graphics,epsfig,verbatim,bm,latexsym,amsmath,url,amsbsy,natbib,multirow,rotating,tikz}
\usepackage[textheight=722pt,textwidth=448pt,centering,headheight=50pt,headsep=12pt,footskip=12pt]{geometry}
\usepackage[colorlinks,citecolor=blue,linkcolor=purple!60!black,urlcolor=purple!60!black]{hyperref}
\usepackage[framemethod=TikZ]{mdframed}

\begin{document}
\begin{center}
\textbf{{\LARGE{When and when not to use optimal model averaging}}}
\end{center}\vspace*{0.5cm}

{\Large
\begin{center}
Michael Schomaker\footnote{University of Cape Town,
Centre for Infectious Disease Epidemiology and Research,
Observatory, 7925; Cape Town, South Africa, \href{mailto:michael.schomaker@uct.ac.za}{michael.schomaker@uct.ac.za}
},
Christian Heumann\footnote{Ludwig-Maximilians Universit{\"a}t M{\"u}nchen, Institut f{\"u}r Statistik, M{\"u}nchen, Germany, \href{mailto:christian.heumann@stat.uni-muenchen.de}{christian.heumann@stat.uni-muenchen.de}}
\end{center}
}

\begin{abstract}
Traditionally model averaging has been viewed as an alternative to model selection with the ultimate goal to incorporate the uncertainty associated with the model selection process in standard errors and confidence intervals by using a weighted combination of candidate models. In recent years, a new class of model averaging estimators has emerged in the literature, suggesting to combine models such that the squared risk, or other risk functions, are minimized. We argue that, contrary to popular belief, these estimators do not necessarily address the challenges induced by model selection uncertainty, but should be regarded as attractive complements for the machine learning and forecasting literature, as well as tools to identify causal parameters. We illustrate our point by means of several targeted simulation studies.
\end{abstract}

\begin{mdframed}[backgroundcolor=red!10, linecolor=black!50]
The published version of this working paper can be cited as follows:\\[0.25cm]
Schomaker, M., Heumann, C.\\
\textit{When and when not to use optimal model averaging}\\
Statistical Papers (2020); 61:2221–2240\\
\url{https://link.springer.com/article/10.1007/s00362-018-1048-3}
\end{mdframed}

\section{Background}
Regression models are the cornerstone of statistical analyses. The motivation for their use is diverse: they might (a) be purely descriptive, (b) target prediction and forecasting problems, (c) help identifying associations or (d) even causal parameters. The motivation for variable selection in regression models is based on the rationale that associational relationships between variables are best understood by reducing the model's dimension. An example would be regression growth models for which a multitude of variables are potentially relevant to describe the relationships in the data (\citealp{salaimartin:2004}). The problem with this approach is that in finite samples (i) the regression parameters after model selection are often biased and (ii) the respective standard errors are too small because they do not reflect the uncertainty related to the model selection process (\citealp{Leeb:2005}; \citealp{Burnham:2002}; \citealp{Hjort:2003}).

A wave of publications in the 1990's (\citealp{Chatfield:1995}; \citealp{Draper:1995}; \citealp{Hoeting:1999}) proposed that the drawback of model selection can be overcome by model averaging. With model averaging one calculates a weighted average of the parameter estimates of a set of candidate models, for example using regression models with a different set of included variables. The weights are determined in such a way that `better' models receive a higher weight. For example, models with a lower AIC may receive a higher weight (\citealp{Buckland:1997}). The variance of these type of estimators is typically calculated such that both the variance related to the parameters of each model and the variance between the different model estimates is taken into account. Note that this approach tackles problem (ii), the incorporation of model selection uncertainty into the standard errors of the regression parameters; but it may not necessarily tackle problem (i) as the regression parameters may still be biased. In fact, model averaging estimators behave similarly to shrinkage estimators because regression coefficients which belong to variables which are not supported among many candidate models are shrunk and are therefore possibly biased. The obvious conclusion is that model averaging is useful to identify associations in regression models and yields more realistic confidence intervals than model selection does. It can therefore serve as a descriptive and exploratory tool in data analysis and be applied in the context of (a) and (c).

However, the pitfalls of this classical model averaging scheme are clear: the estimators produced by a classical weight choice are not optimal from a statistical point of view. The weights are chosen such that one gets improved standard errors. But ideally the weights of an estimator would also result in an averaged estimator which minimizes a risk function, for example the squared risk with respect to some function of $\mu$ (at least asymptotically). This may yield an estimator with good properties, potentially even with good predictive abilities. These type of estimators are known as `Optimal Model Averaging' (OMA) estimators and were mostly inspired by the seminal paper of \citet{Hansen:2007}. He considered a set of nested regression models and proposed to choose weights such that the weighted estimator minimizes a criterion similar to Mallow's $C_p$. With this, the weights are constructed such that the mean squared prediction error is small, therefore one obtains a good bias-variance tradeoff as well as other properties, for example an (asymptotically) optimal estimator based on definitions common in the model averaging literature. The construction of Hansen's estimator corresponds to motivation (b) outlined above. It is no surprise that other authors then also developed optimal model averaging estimators -- based on the same idea, but in the context of different model classes, different loss/risk functions, different model sets, and so on --  see \citealp{Cheng:2015}; \citealp{Gao:2016}; \citealp{Hansen:2008}; \citealp{Hansen:2012}; \citealp{Liang:2011}; \citealp{Liu:2016}; \citealp{Liu:2016b}; \citealp{Zhang:2014}; \citealp{Zhang:2015}; \citealp{Zhang:2016} and the references therein. The interesting part is that the authors of these papers, with few exceptions (e.g. \citealp{Zhang:2015}; \citealp{Zhang:2016}), often motivate their estimators by saying that the purpose for their construction is to overcome the problems of model selection and to include the uncertainty associated with the model selection process. This is surprising as the methodology developed by Hansen and others does not tackle (ii) as needed for (a) and (c), but is rather geared towards (b). Moreover, the construction of confidence intervals is typically not discussed in these papers (but see \citet{Zhang:2017} on interval estimation). Our paper is motivated by this misunderstanding.

We argue that there are at least two different schools of model averaging, each with their own justification and benefit. However, the recent developments in the literature in finding an optimal model averaging estimator should not be confused with the original motivation of `correcting' model selection estimates to include the uncertainty of the model selection process. The motivation of model selection and model averaging originates from the attempts to understand associational structures in models of moderate-to-high dimension (see items (a) and (c)). Optimal model average estimators should rather be seen as additional tools for statistical forecasting and learning problems (see item (b)).

In this paper we are going to demonstrate several points concerning the relationship and differences between different model averaging schemes:
\begin{itemize}
\item we investigate the coverage probability of selected popular model averaging estimators. While recently there has been a moderate interest in understanding the construction of confidence intervals when applying model averaging (\citealp{Kabaila:2016}, \citealp{Wang:2012}, \citealp{Schomaker:2014}, \citealp{Turek:2012}, \citealp{Fletcher:2011}), this topic has been rather under-researched; in particular, it remains unclear how standard model-averaged confidence intervals perform in terms of coverage, and how this compares to naive intervals after model selection.
\item we undertake simulation studies to compare different model averaging approaches under different motivating questions; i.e explanatory, predictive and causal questions of interest.
\item we demonstrate that optimal model averaging can be successfully incorporated into `super learning', a recently proposed data adaptive approach which combines several learners to improve predictive performance.
\item motivated by the above point, we show that OMA can complement procedures which identify causal effects, such as the sequential g-formula. We therefore show that OMA may be of interest even in the context of (d).
\item moreover, we have implemented optimal model averaging estimators in such a way that they can be used easily for super learning and in causal inference.
\end{itemize}
All above points are meant to understand and illustrate under which circumstances the use of optimal model averaging has benefits, and when this is not the case.

\section{Methodological Framework}
Below we review the methods discussed and evaluated in the remainder of this paper. Section \ref{sec:CBMA} introduces criterion based model averaging whereas Sections \ref{sec:MMA}, \ref{sec:JMA}, and \ref{sec:LAE} introduce optimal model averaging estimators. Section \ref{sec:super_learning} describes the concept of super learning. The description of the below methods is brief on purpose, as the contribution of this paper relates to comparison of optimal and traditional model averaging schemes by discussion and simulation.

Consider $n$ observations for which both a response vector $y=(y_1,\ldots,y_n)'$ and a covariate matrix $X=({{x}_1}',\ldots,{{x}_n}')'$, ${x}_i=(x_{i1},\ldots,x_{ip})$, are available. Each variable of $X$ is denoted as ${X}_j=(x_{1j},\ldots,x_{nj})'$. To relate the response with a set of explanatory variables one could use a (regression) model $M_\kappa=f({y}|{X};{\beta})$ for which the parameter vector $\beta$ has to be estimated. If we consider a \textit{set} of candidate models, $\mathcal{M}=\{M_1,\ldots,M_k\}$, for describing ${y}$ based on varying combinations of ${X}_j$'s, then a model selection procedure chooses one single `best' model out of the set $\mathcal{M}$; typically based on some criterion, for example Akaikes Information Criterion (AIC, \citealp{Akaike:1973, Rao:2001}).

\subsection{Criterion Based Model Averaging}\label{sec:CBMA}
With criterion based model averaging, one calculates a weighted average $\hat{\bar{\beta}} = \sum_{\kappa} w_{\kappa} \hat{\beta}_{\kappa}$ from the $k$ estimators $\hat{\beta}_{\kappa}$ ($\kappa=1,\ldots,k$) of the set of candidate (regression) models $\mathcal{M}$ where the weights are calculated in a way such that `better' models receive a higher weight. A popular weight choice would be based on the exponential AIC,
\begin{eqnarray}\label{eqn:weight_AIC}
w_{\kappa}^{\text{AIC}} &=& \frac{\exp(-\frac{1}{2} \mathrm{AIC_{\kappa})}}{\sum_{\kappa=1}^k\exp(-\frac{1}{2} \mathrm{AIC_{\kappa})}}\,,
\end{eqnarray}
where $\mathrm{AIC_{\kappa}}$ is the AIC value related to model $M_{\kappa}\in\mathcal{M}$ (\citealp{Buckland:1997}).  It has been suggested to estimate the variance of the scalar $\hat{\bar{\beta}}_j \in \hat{\bar{{\beta}}}$ via
\begin{eqnarray}\label{formula_MA_var}
\widehat{\text{Var}}(\hat{\bar{\beta}}_j)&=& \left\{\sum_{\kappa=1}^k w_{\kappa} \sqrt{\widehat{\text{Var}}(\hat{\beta}_{j,\kappa}|M_{\kappa}) + (\hat{\beta}_{j,\kappa}-\hat{\bar{\beta}}_j})^2\right\}^2\,,
\end{eqnarray}
where $\hat{\beta}_{j,\kappa}$ is the $j^{th}$ regression coefficient of the $\kappa^{th}$ candidate model. While formula (\ref{formula_MA_var}) from \citet{Buckland:1997} is the most popular choice to calculate standard errors in model averaging, it has also been criticized that the coverage probability of interval estimates based on (\ref{formula_MA_var}) may not always reflect the nominal level (\citealp{Hjort:2003}).

From the Bayesian perspective the quality of a regression model $M_{\kappa} \in \mathcal{M}$ may be judged upon the estimated posterior probability that this model is correct, that is
\begin{eqnarray}\label{eqn:posterior_probability}
\Pr(M_{\kappa}|y) &\propto& \Pr(M_{\kappa}) \int \Pr(y|M_{\kappa},
\beta_{\kappa})\cdot \Pr(\beta_{\kappa}|M_{\kappa}) \,
d\beta_{\kappa} \,,
\end{eqnarray}
where $\Pr(M_{\kappa})$ is the prior probability for the model $M_{\kappa}$ to be correct, $\Pr(y|M_{\kappa}, \beta_{\kappa})$ $=$ $\mathcal{L}({\beta})$ represents the likelihood, and $\Pr(\beta_{\kappa}|M_{\kappa})$ reflects the prior density of $\beta_{\kappa}$ when
$M_{\kappa}$ is the model under consideration. Since, for a large sample size, $\Pr(M_{\kappa}|y)$ can be approximated via the Bayes-Criterion of Schwarz (BCS, BIC), it is often suggested that the weight
\begin{eqnarray}\label{eqn:weight_BIC}
w_{\kappa}^{\text{BIC}} &=& \frac{\exp(-\frac{1}{2} \mathrm{BIC_{\kappa})}}{\sum_{\kappa=1}^k\exp(-\frac{1}{2} \mathrm{BIC_{\kappa})}}\,,
\end{eqnarray}
should be used for the construction of the Bayesian Model Averaging estimator. The BIC corresponds to $-2\mathcal{L}(\hat{\beta}) + p \ln n$, where $p$ corresponds to the number of parameters. The variance of $\hat{\bar{\beta}}_j$ can be estimated in various (similar) ways, depending on the assumptions about the priors and the practical approach of solving the integral in (\ref{eqn:posterior_probability}). Broadly, variance estimation is based on variance decomposition such as the law of total variance, i.e. using
    \begin{eqnarray}\label{eqn:MA_var_Bayes}
\widehat{\text{Var}}(\hat{\bar{\beta}}_j) &=& \widehat{\text{E}}_{\mathcal{M}}(\widehat{\text{Var}}(\hat{\beta}_{j,\kappa}|y,M_{\kappa})) + \widehat{\text{Var}}_{\mathcal{M}}(\widehat{\text{E}}(\hat{\beta}_{j,\kappa}|y,M_{\kappa}))\,,
    \end{eqnarray}
see also \citet{Draper:1995}. Practically, this yields similar, but not identical results as (\ref{formula_MA_var}). Based on the above variance estimates, Bayesian credibility intervals can be constructed.

There are many variations and subtleties when it comes to the implementation of the above estimators. For example, for computational feasibility, one may restrict the number of candidate models. Reviews on Frequentist and Bayesian Model Averaging can be found in \cite{Wang:2009} and \citet{Hoeting:1999}.

\subsection{Mallow's Model Averaging}\label{sec:MMA}
Hansen considers a situation of $k$ nested linear regression models for $k$ variables. Let $\hat{\beta}_{\kappa}$ be the estimated regression parameter of model $M_{\kappa}$, and $\hat{\bar{\beta}}= \sum_{\kappa=1}^k w_{\kappa} \hat{\beta}_{\kappa}$ be a model averaging estimator with $\hat{\mu}_w=X_{k}\hat{\bar{\beta}}$. Based on similar thoughts as in the construction of Mallow's $C_p$ (\citealp{Mallows:1973}), Hansen suggests to minimize the mean squared (prediction) error [MSPE] by minimizing the following criterion:
\begin{eqnarray}\label{eqn:Cp}
\tilde{C}_p &=& (y-X_{k}\hat{\bar{\beta}})'(y-X_{k}\hat{\bar{\beta}})+2{\sigma}^2 K_w\,,
\end{eqnarray}
where $K_w=\text{tr}\,(P_w)$, $P_w=\sum_{\kappa=1}^k w_{\kappa} X_{\kappa}(X_{\kappa}'X_{\kappa})^{-1}X_{\kappa}'$ and $\sigma^2$ is the variance which needs to be estimated from the full model. Consequently, the weight vector $w$ is chosen such that $\tilde{C}_p$ is minimized
\begin{eqnarray}\label{eqn:weights_MMA}
w^{\text{MMA}} &=& \text{arg}\,\min_{_{_{w\in\mathcal{H}}}} \, \tilde{C}_p\,,
\end{eqnarray}
with $\mathcal{H}=\{w=(w_{1},\ldots,w_k) \in [0,1]^k: \sum_{\kappa=1}^k w_{\kappa} =1\}$. Model averaging based on the weight choice (\ref{eqn:weights_MMA}) is often called Mallow's Model Averaging (MMA). MMA has beneficial properties, i.e. it minimizes the MSPE and is asymptotically optimal, see \citet[Theorem 1, Lemma 3]{Hansen:2007} for more details. Moreover, it has been shown that the MMA estimator has a smaller MSE than the OLS estimator (\citealp{Zhang:2016b}).

Since the first part of (\ref{eqn:Cp}) is quadratic in $w_{\kappa}$ and the second one linear, one can obtain the model averaging estimator by means of quadratic programming.

The assumptions of a discrete weight set and nested regression models sound restrictive, but it has been shown that both assumptions are not necessarily required and MMA can be applied to non-nested regression models as well; given that this is computationally feasible (\citealp{Wan:2010}).

\subsection{Jackknife Model Averaging}\label{sec:JMA}
Jacknife Model Averaging (JMA) as proposed by \citet{Hansen:2012} for linear models, builds on leave-one-out (LOO)  cross validation. For Model $M_{\kappa}$ the LOO residual vector is $\tilde{\epsilon}^{\kappa}=y-\hat{y}^{\kappa}$, with $\hat{y}^{\kappa}=x_i^{\kappa}(X_{(-i)}^{\kappa'}X_{(-i)}^{\kappa})^{-1}X_{(-i)}^{\kappa'}y_{(-i)}$ where the index $_{(-i)}$ describes that the respective matrix excludes observation $i$, $i=1,\ldots,n$. It can be shown that there is a simple algebraic relationship which allows the computation of the LOO residuals in one rather than $n$ operations:
\begin{eqnarray}
\tilde{\epsilon}^{\kappa} = D_{\kappa}\hat{\epsilon}^{\kappa}
\end{eqnarray}
where $\hat{\epsilon}^{\kappa}$ is the standard least squares residual vector $y-P_{\kappa}y$ with the hat matrix $P=X(X'X)^{-1}X'$; and $D_{\kappa}$ is a $n \times n$ diagonal matrix with $D_{ii,\kappa}=(1-P_{ii,\kappa})^{-1}$, $i=1,...,n$.

For $k$ candidate models the linear weighted LOO residuals are $\tilde{\epsilon}_w=\sum_{\kappa} w_{\kappa}\tilde{\epsilon}^{\kappa}$, $\kappa=1,\ldots,k$. An estimate of the true expected squared error is $\text{CV}_w = n^{-1}\tilde{\epsilon}_w'\tilde{\epsilon}_w$ and an appropriate weight choice would thus be
\begin{eqnarray}\label{eqn:weights_JMA}
w^{\text{JMA}} &=& \text{arg}\,\min_{_{_{w\in\mathcal{H}}}} \, \text{CV}_w\,,
\end{eqnarray}

As with MMA, the weights can be obtained with quadratic programming. The estimator has similar properties as the MMA estimator (\citealp{Hansen:2012}). Model averaging with the weight choice (\ref{eqn:weights_JMA}) is called Jackknife Model Averaging.

\subsection{Lasso Averaging}\label{sec:LAE}
Shrinkage estimation, for example via the LASSO (\citealp{Tibshirani:1996}), can be used for model selection. This requires the choice of a tuning parameter which comes with tuning parameter selection uncertainty. LASSO averaging estimation (LAE), or more general shrinkage averaging estimation (\citealp{Schomaker:2012}), is a way to combine shrinkage estimators with different tuning parameters.

Consider the LASSO estimator for a simple linear model:
\begin{eqnarray}\label{eqn:lasso}
\hat{\beta}_{\text{LE}}({\lambda}) &=& \text{arg}\,\min\,\left\{\sum_{i=1}^n (y_i - \beta_0 - \sum_{j=1}^p x_{ij}{{\beta}}_j)^2 + {\lambda} \sum_{j=1}^p |\beta_j| \right\}\,.
\end{eqnarray}
The complexity parameter $\lambda \geq 0$ tunes the amount of shrinkage and is typically estimated via the generalized cross validation criterion or any other cross validation criterion. The larger the value of $\lambda$, the greater the amount of shrinkage since the estimated coefficients are shrunk towards zero.

Consider a sequence of candidate tuning parameters ${\lambda}=\{\lambda_1,\ldots,\lambda_L\}$. If each estimator $\hat{\beta}_{\text{LE}}({\lambda_i})$ obtains a specific weight $w_{\lambda_i}$, then a {\textit{LASSO averaging estimator}} takes the form
\begin{eqnarray}\label{eqn:LAE}
\hat{\bar{\beta}}_{\text{LAE}}&=& \sum_{i=1}^L w_{\lambda_i} \hat{\beta}_{\text{LE}}({\lambda_i}) = {w}_\lambda \hat{{B}}_{\text{LE}}  \,,
\end{eqnarray}
where $\lambda_i \in [0,c]$, $c>0$ is a suitable constant, $\hat{{{B}}}_{\text{LE}}=(\hat{\beta}_{\text{LE}}(\lambda_1),\ldots,\hat{\beta}_{\text{LE}}(\lambda_L))'$ is the $L \times p$ matrix of the LASSO estimators, ${w}_\lambda=(w_{\lambda_1},\ldots,w_{\lambda_L})$ is an $1 \times L$ weight vector, ${w}_\lambda \in \mathcal{W}$ and  $\mathcal{W}=\{{w}_\lambda \in [0,1]^L: {1}'{w}_\lambda=1\}$.

A general measure for the cross validation error with squared loss function would be
\begin{eqnarray}\label{eqn:OCV}
OCV_k &=& n^{-1}{\tilde{\epsilon}_\kappa(w)}' \tilde{\epsilon}_\kappa(w)  \propto {w}_{\lambda} {E}'_k {E}_k {{w}_{\lambda}}' \,,
\end{eqnarray}
where ${E}_k = (\tilde{\epsilon}_k(\lambda_1),\ldots,\tilde{\epsilon}_k(\lambda_L))$ is the $n \times L$ matrix of the $k$-fold cross-validation residuals for the $L$ competing tuning parameters. An optimal weight vector for this criterion is then
\begin{eqnarray}\label{eqn:weights_OCV}
w^{\text{LAE}} &=& \text{arg}\,\min_{_{_{w\in\mathcal{W}}}} OCV_k \,.
\end{eqnarray}

These weights can also be calculated with quadratic programming. More details can be found in \citet{Schomaker:2012}.

\subsection{Super Learning}\label{sec:super_learning}
Depending on the specific problem, optimal model averaging as described in the above sections may be a good prediction algorithm or not. To choose and combine the best prediction methods, \textit{super learning}\index{super learner} can be used. Super learning means considering a set of prediction algorithms, for example regression models, shrinkage estimators or model averaging. Instead of choosing the algorithm with the smallest cross validation error, super learning chooses a weighted combination of different algorithms, that is the weighted combination which minimizes the cross validation error. It can be shown that this weighted combination will perform (asymptotically) at least as good as the best algorithm, if not better (\citealp{vanderLaan:2008}) and is known as the oracle property of super learning.

For example: consider Learner 1 (L1) to be a linear model including all available covariates and learner 2 (L2) to be Mallow's Model Averaging. Both of them have a specific $k$-fold cross-validation error, for a given loss function, that is $\text{CV}_{k}^{L1}$ and $\text{CV}_{k}^{L2}$. Now, find the linear combination of the two predictions from L1 and L2 that best predicts the outcome. This can be achieved by non-negative least squares estimation, as (for the above mentioned oracle property to hold) the weights need to be positive and sum up to one. The final prediction algorithm is then the weighted linear combination of the two learners. The cross validation error of this combination is then asymptotically at least as low (and therefore good) as the errors  $\text{CV}_{k}^{L1}$ and $\text{CV}_{k}^{L2}$.

The interested reader is referred to \citet{vanderLaan:2007} and \citet{vanderLaan:2011}, and the references therein, for more details.

\section{Simulation Studies}
The purpose of this section is to contrast simple traditional model averaging, both frequentist and bayesian as described in Section \ref{sec:CBMA} with optimal model averaging as described in Sections \ref{sec:MMA}, \ref{sec:JMA}, and \ref{sec:LAE} for different situations. The first setting described in Section \ref{sec:sim_linear} targets linear regression settings motivated by (a) and (c), i.e. those where regression is meant to describe associational relationships. The next Section \ref{sec:sim_prediction} targets (b), that is the use of regression for prediction. Finally, in Section \ref{sec:sim_causal}, we look at longitudinal data for which (d), i.e. the identification of a causal effect, is of interest.

\subsection{Associations in a Linear Regression Model}\label{sec:sim_linear}
In this setup, we compare different estimators of a linear regression model: the ordinary least squares estimate of the full model [OLS], the model selection estimates of the model selected by AIC [MS], traditional model averaging estimates based on the weight choices (\ref{eqn:weight_AIC}) [FMA] and (\ref{eqn:weight_BIC}) [BMA], and Mallow's model averaging estimates based on the weight choice (\ref{eqn:weights_MMA}) [MMA]. We selected the above estimators because they reflect the most popular approaches in the literature. Additionally, for BMA, we follow the implementation from the $R$-package \texttt{BMA} (\citealp{Raftery:2017}), which uses a subset of candidate models based on a leaps and bounds algorithm in conjunction with ``Occam's razor'', see \citet{Hoeting:1999} for more details. Frequentist model averaging is based on all possible candidate models, model selection is based on those models selected by stepwise selection with AIC, and optimal model averaging on the set of nested models. Variance estimates for FMA and MMA are based on (\ref{formula_MA_var}), and those of BMA are based on (\ref{eqn:MA_var_Bayes}). Confidence intervals were constructed using critical values from a standard normal distribution, as often done in naive regression analyses.

The setup of our simulation is as follows: We generate 10 variables (sample size: $n=500$) using normal, log-normal and exponential distributions: $\mathbf{X}_1, \mathbf{X}_2, \mathbf{X}_3, \mathbf{X}_4 \sim \text{N}(0,1)$, $\mathbf{X}_5, \mathbf{X}_6, \mathbf{X}_7 \sim \text{logN}(0,0.5)$, $\mathbf{X}_8, \mathbf{X}_9, \mathbf{X}_{10} \sim \text{Exp}(1)$. To model the dependency between the covariates we use a Clayton Copula (\citealp{Yan:2007}) with a copula parameter of $1$ which indicates moderate correlation among the covariates. We then define $\mu_y= 1X_3 + 2X_4 + 3X_5 + 3X_6 + 2X_7 +1X_8 + 0.5X_9$ and generate the outcome from $N(\mu_y,\exp(2))$. Therefore, 7 out of 10 variables have an effect of different size on $y$.

We compare the point estimates of the five approaches in terms of unbiasedness. Secondly, we compare estimated standard errors for model averaging estimators) with those obtained from the simulation study (i.e. based on the variance of $\hat{\bar{\beta}}$ over the $R=5000$ simulation runs). Thirdly, we evaluate the coverage probability of the respective 95\% interval estimates.

\subsection{Forecasting}\label{sec:sim_prediction}
This setup targets prediction accuracy. We generate 10 variables (sample size: $n=500$) using again normal, log-normal and exponential distributions: $\mathbf{X}_1, \mathbf{X}_2, \mathbf{X}_3, \mathbf{X}_4 \sim \text{N}(0,1)$, $\mathbf{X}_5, \mathbf{X}_6, \mathbf{X}_7 \sim \text{logN}(0,0.5)$, $\mathbf{X}_8, \mathbf{X}_9, \mathbf{X}_{10} \sim \text{Exp}(1)$. To model the dependency between the covariates we again use a Clayton Copula with a copula parameter of $1$. We then define $\mu_y= -5 + 0.5X_2 + 1.5X_6 + 1.5X_9 +X_6\times X_9 + X_2^2$ and generate the outcome from $N(\mu_y,\exp(1.5))$. Therefore, 3 out of 10 variables predict $y$ and both interactions and non-linear associations are present.

We evaluate the mean squared prediction error for the same methods evaluated in Section \ref{sec:sim_linear}, i.e. OLS, MS, FMA, BMA, and MMA. In addition we evaluate the predictive performance of super learning with two different types of learner sets: the first one  (SL) consists of the OLS of the full linear model, random forests (\citealp{Breiman:2001}), stepwise regression based on AIC, the LASSO, the arithmetic mean,  GLM's based on EM-algorithm-Bayesian model fitting (\citealp{Gelman:2016}), additive models (\citealp{Wood:2006}), and the full linear model with interactions, with and without model selection with AIC. The second learner set (SL+) consists of all learners from the first set, but adds Jacknife Model Averaging, Lasso Averaging and Mallows Model Averaging to the learner set. All of these estimators are fitted i) with the full set of variables, ii) with the full set plus all two-way interactions and iii) with the full set plus squared transformations of all variables. For Lasso Averaging we used a $\lambda$-sequence of length $100$, where the maximum $\lambda$-value is the smallest value for which all coefficients are zero, the minimum $\lambda$ value is $0.0001$, and all other $\lambda$-values are equally spaced between these two (on a log-scale). While this is a common approach (\citealp{Friedman:2010}), alternative sequences can be easily specified in common software packages, such as the $R$-package \texttt{glmnet}.

In this simulation, both the mean squared prediction error as well as the choice of learners from the super learner algorithm are of interest. The simulation is based on $5000$ runs.

\subsection{Causal Inference}\label{sec:sim_causal}
This simulation is inspired by the HIV treatment analyses of \cite{Schomaker:2018} and \cite{Schomaker:2016}. We generate longitudinal data ($t=0,1,\ldots,6$) for 3 time-dependent confounders ($\mathbf{L_t}=\{L^1_t,L^2_t,L^3_t\}$), an outcome $(Y_t)$, an intervention $(A_t)$, as well as baseline data at $t=0$ for 7 variables, using structural equation models (\citealp{Sofrygin:2016}). The data generating mechanism  is described in detail in Appendix \ref{sec:appendix_data_generating}. In this simulation we are interested in the expected counterfactual outcome at the end of follow-up (i.e. $t=6$) which would have been observed under 2 different intervention rules $\bar{d}^j$, $j=1,2$, which assign treatment ($A_t$) either always (at each time point) or depending on the confounders, i.e. $A_t=1$ if $L^1_t <350$ or $L^2_t < 15\%$ or $L^3_t < -2$; that is we want to estimate $E(Y_6^{\bar{d}_j})$ [see Appendix \ref{sec:appendix_data_generating} for more details regarding notation]. We denote these target quantities as $\psi_1$ and $\psi_2$ and their true values are $-1.80$ and $-2.02$ respectively. They can be estimated using appropriate methodology, for example using the sequential g-formula; see Appendix \ref{sec:appendix_g_formula} for more details. Briefly, for each point in time, i.e. $t=6,\ldots,1,0$, the conditional outcome given the covariate history needs to be estimated. To avoid model mis-specification, it is common to use super learning for this. We use super learning with two different sets of learners. The first one consists of the OLS of the full linear model, the arithmetic mean, stepwise regression based on AIC, GLM's based on EM-algorithm-Bayesian model fitting, additive models, and linear models with interactions. The second learner set consists of all learners from the first set, but adds Jacknife Model Averaging, Lasso Averaging (as specified in Section \ref{sec:sim_prediction}) and Mallows Model Averaging to the learner set. All of these estimators are fitted i) with the full set of variables, ii) with the full set plus all two-way interactions and iii) with the full set plus squared transformations of all variables. The simulation is based on $1000$ runs.

This simulation compares bias and coverage with respect to the two different learners and interventions respectively; moreover, we are particularly interested whether super learning, applied in a complex longitudinal setup, picks optimal model averaging estimators for the fitting process or not. This point is not immediately clear as simple learners, such as additive models and GLM's with interaction, are already complex enough to model the data-generating process described in Appendix \ref{sec:appendix_data_generating}. Whether a weighted combination including OMA is of benefit is the motivation of this simulation.

\subsection{Results}
The results of the first simulation study are summarized in Table \ref{tab:sim}.

\begin{table}[ht!]
\caption{\label{tab:sim} Results of the first simulation study. `est' refers to the estimated standard error of the respective method, averaged over the simulation runs; `sim' refers to the simulated standard error based on the variation of the point estimates over the simulation runs. \vspace*{0.25cm}}
\centering
{\small{
\fbox{%
\begin{tabular}{r|p{0.0425\textwidth}p{0.0425\textwidth}p{0.0425\textwidth}p{0.0425\textwidth}p{0.0425\textwidth}p{0.0425\textwidth}p{0.0425\textwidth}p{0.0425\textwidth}p{0.0425\textwidth}p{0.0425\textwidth}}
\multicolumn{1}{l}{(a)}& \multicolumn{10}{c}{Point Estimates}\\
\hline
Method&$\beta_1$&$\beta_2$&$\beta_3$&$\beta_4$&$\beta_5$&$\beta_6$&$\beta_7$&$\beta_8$&$\beta_9$&$\beta_{10}$\\
\hline
  OLS &  0.00 & 0.00 & 1.00 & 2.01 & 3.00 & 3.00 & 1.98 & 1.00 & 0.49 & 0.00 \\
   MS &  0.02 & 0.02 & 0.95 & 2.04 & 3.04 & 3.03 & 1.98 & 0.97 & 0.38 & 0.02 \\
  FMA &  0.04 & 0.04 & 0.90 & 2.06 & 3.06 & 3.06 & 1.93 & 0.93 & 0.37 & 0.02 \\
  BMA &  0.04 & 0.03 & 0.77 & 2.18 & 3.18 & 3.17 & 1.85 & 0.82 & 0.22 & 0.02 \\
  MMA &  0.10 & 0.06 & 1.03 & 2.01 & 2.97 & 2.93 & 1.85 & 0.86 & 0.31 & 0.00 \\
\hline
\multicolumn{1}{c}{}&&&&&&&&&&\\
\multicolumn{1}{l}{(b)} & \multicolumn{10}{c}{Standard Errors}\\
\hline
Method&$\beta_1$&$\beta_2$&$\beta_3$&$\beta_4$&$\beta_5$&$\beta_6$&$\beta_7$&$\beta_8$&$\beta_9$&$\beta_{10}$\\
\hline
OLS -- est   &0.44  &0.44  &0.44  &0.44  &0.66  &0.66  &0.66  &0.39  &0.39  &0.39 \\
OLS -- sim   &(0.44)  &(0.43)  &(0.44)  &(0.44)  &(0.66)  &(0.65)  &(0.67)  &(0.39)  &(0.38)  &(0.39) \\
MS  -- est   &0.07  &0.07  &0.35  &0.42  &0.64  &0.64  &0.61  &0.34  &0.18  &0.05 \\
MS  -- sim   &(0.34)  &(0.33)  &(0.53)  &(0.44)  &(0.66)  &(0.66)  &(0.75)  &(0.46)  &(0.44)  &(0.30) \\
FMA -- est   &0.32  &0.32  &0.47  &0.44  &0.67  &0.67  &0.71  &0.42  &0.35  &0.28 \\
FMA  -- sim  &(0.28)  &(0.26)  &(0.52)  &(0.45)  &(0.68) & (0.68) &(0.79)  &(0.47)  &(0.37)  &(0.25) \\
BMA -- est   & 0.13 & 0.12 &0.45  &0.44  &0.67  &0.67  &0.73  &0.41  &0.25  &0.11 \\
BMA -- sim   &(0.17)  &(0.15)  &(0.63)  &(0.48)  &(0.74)  &(0.74)  &(1.00)  &(0.58)  &(0.36)  &(0.13) \\
MMA -- est   &0.50  &0.45  &0.46  &0.48  &0.73  &0.75  &0.76  &0.44  &0.26  &0.09 \\
MMA -- sim   &(0.44)  &(0.43)  &(0.43)  &(0.44)  &(0.66)  &(0.67)  &(0.69)  &(0.42)  &(0.34)  &(0.20) \\
 \hline
\multicolumn{1}{c}{}&&&&&&&&&&\\
\multicolumn{1}{l}{(c)} & \multicolumn{10}{c}{Coverage Probability (in \%)}\\
\hline
Method&$\beta_1$&$\beta_2$&$\beta_3$&$\beta_4$&$\beta_5$&$\beta_6$&$\beta_7$&$\beta_8$&$\beta_9$&$\beta_{10}$\\
\hline
OLS &  94 & 96 & 95 & 95 & 95 & 95 &95 & 95 & 95 & 95 \\
 MS &  94 & 94 & 80 & 94 & 94 & 95 &92 & 86 & 44 & 94 \\
FMA &  99 & 99 & 85 & 95 & 95 & 95 &90 & 87 & 79 & 99 \\
BMA &  99 & 100 & 67 & 92 & 93 & 93 &81 & 73 & 44 & 100 \\
MMA &  97 & 97 & 96 & 96 & 97 & 97 &95 & 90 & 60 & 100 \\
\hline
\multicolumn{1}{c}{}&&&&&&&&&&\\
\multicolumn{1}{l}{(d)} & \multicolumn{10}{c}{MSE with respect to $\mu_y$}\\
\hline
Method&OLS&MS&FMA&BMA&MMA&&&&&\\
\hline
MSE & 1.214  & 1.261  & 1.197  & 1.526  & 1.213  &  & &  &  & \\
\hline
\end{tabular}
}}}
\end{table}

\clearpage
As expected the OLS is approximately unbiased, whereas the other estimators are not necessarily unbiased, particularly around the small effects of $\beta_3$, $\beta_8$ and $\beta_{9}$ (Table 1a). This simple but important property is often neglected in the model averaging literature. One reason might be that optimality in the model selection literature is typically defined to be either consistency (choosing, asymptotically, the correct model out of a set of candidate models -- given that the candidate model is contained in the set) or efficiency (the selected model minimizes, asymptotically, a risk function -- based on the assumption of a true model of infinite dimension). See \citet{Leeb:2008} for more details.

Table 1b contrasts the average estimated standard errors with those obtained from the simulations, i.e. the variance of the point estimates over the 5000 simulation runs. Ideally they should be as close as possible. It can be seen that the estimated standard errors are appropriate for the OLS estimator, and too small for the model selection estimator. This highlights the problems of model selection uncertainty. Model averaging by means of using AIC weights performs much better, addressing the issues related to model selection, but there is a tendency towards over-conservativeness, rather than over-confidence. Bayesian Model Averaging, with a restricted set of candidate models based on  the approach explained earlier, produces less variability and doesn't seem to produce very accurate standard errors, though they are still somewhat superior to model selection. MMA obviously doesn't perform very well when it comes to estimating the standard errors of $\beta_9$ and $\beta_{10}$; this is because of its nested model setup, but also because the approach of using (\ref{formula_MA_var}) for variance estimation is rather pragmatic. As highlighted before, MMA has been developed for point estimation. However, where computationally feasible, the performance of MMA can potentially be improved by using the set of all (i.e. non-nested) models, and possibly by also using bootstrapping for confidence interval estimation. In our setting, this improved coverage for those variables which were specified to be in a rear position, i.e. coverage for $\beta_9$ improved from 60\% to 88\% and coverage of $\beta_8$ improved from 90\% to 91\%; but often the quadratic programming problem required minor modifications of the first part of (\ref{eqn:Cp}) to be solvable.

A look at the coverage probabilities reveals the problems of both model selection and model averaging: particularly for the small effects the actual coverage is way below the nominal coverage. This is not necessarily surprising because the distribution of model averaging estimators can be non-normal (\citealp{Hjort:2003}). To solve this problem re-sampling may be a viable option (\citealp{Schomaker:2014}), though there are valid theoretical concerns around this as well (\citealp{Poetscher:2006}). Alternatively, one may simply use the OLS interval estimates of the full model as they are asymptotically equivalent to the estimator from \citet{Hjort:2003}, see \citet{Wang:2012} for more details.

The results of the second simulation study are summarized in Table \ref{tab:sim2}.

\begin{table}[ht!]
\caption{\label{tab:sim2}Results of the second simulation study. (a) estimated mean squared prediction error, with standard error, for different model selection and model averaging techniques. Prediction with super learning contains both a set of learners with optimal model averaging techniques (SL+) and without (SL). (b) the weight for each learner, averaged over the simulaton runs, is listed as well.   \vspace*{0.25cm}}
\centering
{\small{
\fbox{%
\begin{tabular}{r|p{0.08\textwidth}p{0.08\textwidth}p{0.08\textwidth}p{0.08\textwidth}p{0.08\textwidth}p{0.08\textwidth}p{0.08\textwidth}}
\multicolumn{1}{l}{(a)} & \multicolumn{7}{c}{Predictive Performance}\\
\hline
 & OLS & MS & FMA & BMA & MMA & SL & SL+ \\
  \hline
MSPE & 22.38 & 22.34 & 31.72 & 22.36 & 22.39 & 21.47 & 21.12 \\
s.e. &  1.45 & 1.45 & 3.09 & 1.46 & 1.45 & 1.41 & 1.39 \\
   \hline
\multicolumn{1}{l}{}&&&&&&&\\
\multicolumn{1}{l}{(b)} & \multicolumn{7}{c}{Choice of Learners}\\
\hline
&learner & weight & learner & weight & learner & weight&\\
\hline
&    MMA               &0.0002&LAE              &0.0022&JMA              &0.0002&\\
&    MMA (+Int.)       &0.0038&LAE  (+Int.)     &0.1032&JMA (+Int.)      &0.0044&\\
&    MMA (+squ.)       &0.1588&LAE  (+squ.)     &0.3405&JMA (+squ.)      &0.1588&\\
&GLM (Bayes)&0.0000&GLM (+AIC) &0.0366&GLM (+Int.)       &0.0174&\\
&random forest     &0.0357&LASSO      &0.0024&GLM (+AIC/Int.)   &0.0870&\\
&mean       &0.0001&GLM        &0.0001&GAM        &0.0138&\\
\end{tabular}
}}}
\end{table}

It can be clearly  seen that model averaging and model selection can't improve the mean squared prediction error in this setting. However, super learning provides much better predictive accuracy. In particular, super learning using optimal model averaging (SL+) has the best overall performance.

In the second simulation the most heavily utilized learners are Lasso averaging including squared variables, as well as JMA and MMA with transformations. As expected, optimal model averaging can help to improve predictive accuracy, in particular when used in conjunction with super learning.

\begin{figure}[ht!]
\begin{center}
\includegraphics[scale=0.55]{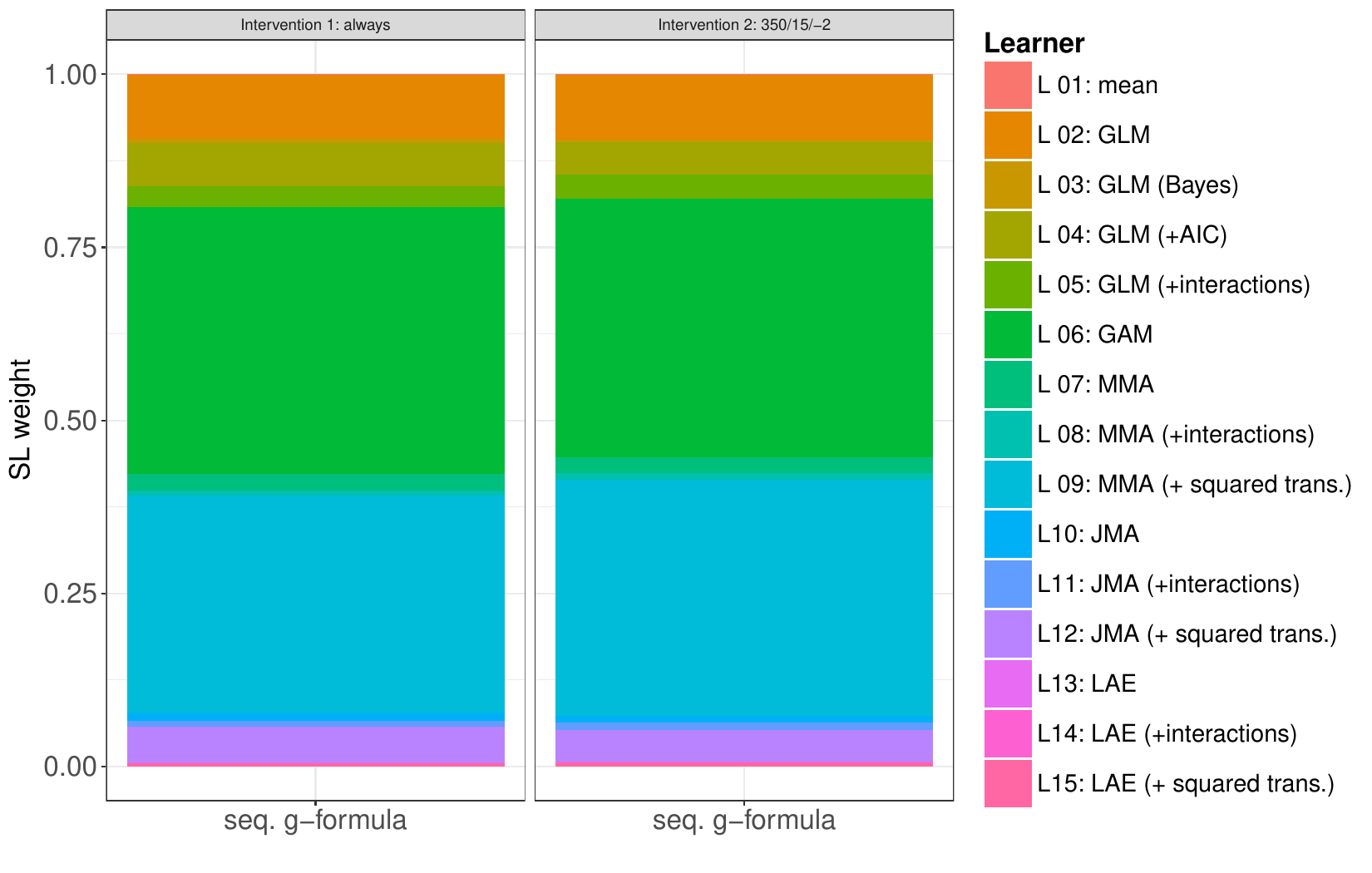}
\caption{Results of the third simulation study: the weight for each learner, averaged over the simulation runs.}
\label{figure:learners}
\end{center}
\end{figure}

The results of the third simulation study are summarized in Table \ref{tab:sim3} and Figure \ref{figure:learners}. It can be seen that in a complex longitudinal setup, with 6 follow-up times, and a data-generating process which includes non-linearities and interactions, a couple of learners contribute most to the estimation process; that is, additive models, MMA with squared transformations and JMA with squared transformations, as well as simple GLM's. This implies that even when learners are available which already describe the data-generating process well (here: GAM's and GLM's with interactions), optimal model averaging can still be utilized by super learning and thus be of benefit.

\begin{table}[ht!]
\caption{\label{tab:sim3} Results of the third simulation study: bias and coverage for different sets of learners and different interventions. \vspace*{0.25cm}}
\centering
{\small{
\fbox{%
\begin{tabular}{llll}
Intervention & Learner Set & Bias & Coverage\\
\hline
always       & without OMA & 0.036 & 90\%\\
always       & with OMA    & 0.036 & 91\%\\
350/15\%/-2  & without OMA & 0.12  & 97\%\\
350/15\%/-2  & with OMA    & 0.10  & 97\%\\
\end{tabular}
}}}
\end{table}

Model mis-specification is a crucial concern when identifying causal parameters (\cite{vanderLaan:2011}) and this is the motivation for using super learning in this context. In our example, bias after estimation still exists, for both interventions of interest (Table \ref{tab:sim3}). Using optimal model averaging has only a small benefit in terms of reducing bias in this particular setting.

\section{Conclusion}
Model averaging in its traditional sense addresses the problem of model selection uncertainty. Because model averaging can still yield biased estimates and imperfect coverage, its main benefit is in identifying associations in a moderate-to-large data set. Such a procedure can also be helpful in an explorative data analysis. However, these estimators wouldn't necessarily be the first choice for quantifying associations as exactly as possible, for complex prediction problems, or for estimation procedures which seek to identify causal parameters.

In contrast, optimal model averaging as proposed in the recent years may not be ideal to take into account model selection uncertainty as their construction principle is not based on interval estimation. However, the idea of optimal model averaging is attractive in analyses which deal with prediction and forecasting problems. Some of these estimators, such like Mallow's Model Averaging, are computationally efficient, robust, and tackle predictions from a different angle. This may benefit existing approaches, such as super learning, where a broad spectrum of learners are required. Super learning techniques are a popular tool in the process of identifying a causal quantity of interested by means of targeted maximum likelihood estimation (\citealp{Gruber:2012}, \citealp{Petersen:2014}). Therefore, the benefit of optimal model averaging techniques may reach far beyond pure prediction problems and play its role in causal analyses.

Our recommendation is that future manuscripts that propose optimal model averaging techniques focus their motivation and data examples around prediction (or the use of prediction in estimating causal quantities) rather than model selection uncertainty questions.

\appendix
{\small

\section{Software}
We have implemented Mallow's Model Averaging, Jackknife Model Averaging, and Lasso Averaging in the $R$-package MAMI (\citealp{Schomaker:2017}), available at \url{http://mami.r-forge.r-project.org/}. In addition to this, we have implemented several wrappers that make optimal model averaging easily useable for super learning (\citealp{Polley:2017}), and in conjunction with causal inference packages such as \texttt{tmle} (\citealp{Gruber:2012}) and \texttt{ltmle} (\citealp{Lendle:2017}). Available wrappers are explained by calling \texttt{listSLWrappers()}, and examples are given in the documentation (\citealp{Schomaker:2017b}).

\section{Notation and Background on the Sequential g-formula}\label{sec:appendix_g_formula}

Consider a sample of size $n$ of which measurements are available both at baseline ($t=0$) and during a series of follow-up times $t=1,\ldots,T$. At each point in time we measure the outcome $Y_t$, the intervention $A_t$, time-dependent covariates $\mathbf{L}_t=\{L^1_t,\ldots,L^q_t\}$, and a censoring indicator $C_t$. $\mathbf{{L}}_{t}$ may include baseline variables $\mathbf{V}=\{L^1_0,\ldots,L^{q_V}_0\}$ and can potentially contain variables which refer to the outcome variable before time $t$, for instance ${Y}_{t-1}$. The treatment and covariate history of an individual $i$ up to and including time $t$ is represented as $\bar{A}_{t,i}=(A_{0,i},\ldots,A_{t,i})$ and $\bar{L}^s_{t,i}=(L^s_{0,i},\ldots,L^s_{t,i})$ respectively. $C_t$ equals $1$ if a subject gets censored in the interval $(t-1,t]$, and $0$ otherwise. Therefore, $\bar{C}_t=0$ is the event that an individual remains uncensored until time $t$.

The counterfactual outcome $Y_{t,i}^{\bar{a}_{t}}$ refers to the hypothetical outcome that would have been observed at time $t$ if subject $i$ had received, possibly contrary to the fact, the treatment history $\bar{A}_{t,i}=\bar{a}_t$. Similarly, $\mathbf{L}_{t,i}^{\bar{a}_{t}}$ are the counterfactual covariates related to the intervention $\bar{A}_{t,i}=\bar{a}_t$. The above notation refers to \textit{static} treatment rules; a treatment rule may, however, depend on covariates, and in this case it is called \textit{dynamic}. A dynamic rule ${d}_t(\mathbf{\bar{L}}_{t})$ assigns treatment ${A}_{t,i} \in \{0,1\}$ as a function of the covariate history $\mathbf{\bar{L}}_{t,i}$. The vector of decisions $d_t$, $t=0,...,T$, is denoted as $\bar{d}_T=\bar{d}$. The notation $\bar{A}_t = \bar{d}$ refers to the treatment history according to the rule $\bar{d}$. The counterfactual outcome related to a dynamic rule $\bar{d}$ is $Y_{t,i}^{\bar{d}}$, and the counterfactual covariates are $\mathbf{L}_{t,i}^{\bar{d}}$.

In Section \ref{sec:sim_causal} we consider the expected value of $Y$ at time $6$, under no censoring, for a given treatment rule $\bar{d}$ to be the main quantity of interest, that is $\psi=\mathbb{E}(Y_6^{\bar{d}})$.

The sequential g-formula can estimate this target quantity by sequentially marginalizing the distribution with respect to $\mathbf{L}$ given the intervention rule of interest. It holds that
\begin{eqnarray*}\label{eqn:seq_g_formula}
\mathbb{E}(Y_T^{\bar{d}}) &=& \mathbb{E}(\,\mathbb{E}(\,\ldots\mathbb{E}(\,\mathbb{E}(Y_T|\bar{A}_T=\bar{d}_{T}, \mathbf{\bar{L}}_T) | \bar{A}_{T-1}=\bar{d}_{T-1}, \mathbf{\bar{L}}_{T-1}\,)\ldots|\bar{A}_{0}={d}_{0}, \mathbf{{L}}_{0}\,)|\mathbf{{L}}_{0}\,)\,,
\end{eqnarray*}
see for example \citet{Bang:2005}. Equation (\ref{eqn:seq_g_formula}) is valid under several assumptions: sequential conditional exchangeability, consistency, positivity and the time ordering $\mathbf{L}_t \rightarrow A_t \rightarrow C_t \rightarrow Y_t$. These assumptions essentially mean that all confounders need to be measured, that the intervention is well-defined and that individuals have a positive probability of continuing to receive treatment according to the assigned treatment rule, given that they have done so thus far and irrespective of the covariate history; see \citet{Daniel:2013} and \citet{Robins:2009} for more details and interpretations. Note that the two interventions defined in Section \ref{tab:sim3} also assign $C_t=0$ meaning that we are interested in the effect estimate under no censoring.

To estimate $\psi$ one needs to the the following for $t=T,...,0$: (i) use an appropriate model to estimate $\mathbb{E}(Y_T|{\bar{A}}_{T-1}={\bar{d}}_{T-1}, \mathbf{\bar{L}}_T)$. The model is fit on all subjects that are uncensored (until $T-1$). Note that the outcome refers to the measured outcome for $t=T$ and to the prediction (of the conditional outcome) from step (ii) if $t<T$. Then, (ii) plug in $\bar{A}_t=\bar{d}_{t}$ to predict $Y_t$ at time $t$; (iii) For $t=0$ the estimate $\hat{\psi}$ is obtained by calculating the arithmetic mean of the predicted outcome from the second step.

\section{Data Generating Process in the Simulation Study}\label{sec:appendix_data_generating}

Both baseline data ($t=0$) and follow-up data ($t=1,\ldots,12$) were created using structural equations using the $R$-package \texttt{simcausal} (\citealp{Sofrygin:2016}). The below listed distributions, listed in temporal order, describe the data-generating process motivated by the analysis from \citet{Schomaker:2016}. Baseline data refers to region, sex, age, CD4 count, CD4\%, WAZ and HAZ respectively ($V^1$, $V^2$, $V^3$, $L^1_0$, $L^2_0$, $L^3_0$, $Y_0$), see \citet{Schomaker:2016} for a full explanation of variables and motivational question. Follow-up data refers to CD4 count, CD4\%, WAZ and HAZ ($L^1_t$, $L^2_t$, $L^3_t$, $Y_t$), as well as a treatment ($A_t$) and censoring ($C_t$) indicator. In addition to Bernoulli ($B$), uniform ($U$) and normal ($N$) distributions, we also use truncated normal distributions which are denoted by $N_{[a,b]}$ where $a$ and $b$ are the truncation levels. Values which are smaller $a$ are replaced by a random draw from a $U(a_1,a_2)$ distribution and values greater than $b$ are drawn from a $U(b_1,b_2)$ distribution. Values for $(a_1, a_2, b_1, b_2)$ are $(0,50,5000,10000)$ for $L^1$, (0.03,0.09,0.7,0.8) for $L^2$, and $(-10,3,3,10)$ for both $L^3$ and $Y$. The notation $\bar{\mathcal{D}}$ means ``conditional on the data that has already been measured (generated) according the the time ordering''.\\

{\footnotesize{
\noindent For $t=0$:
\indent\begin{eqnarray*}
V^1 &\sim& B(p=4392/5826) \\
V^2|\bar{\mathcal{D}} &\sim& \left\{ \begin{array}{cl}
               B(p=2222/4392) & \quad \mbox{if} \quad V^1 = 1\\
               B(p=758/1434)  & \quad \mbox{if} \quad V^1 = 0\\
               \end{array}
               \right. \\
V^3|\bar{\mathcal{D}} &\sim& U(1,5) \\
L^1_0|\bar{\mathcal{D}} &\sim & \left\{ \begin{array}{cl}
               N_{[0,10000]}(650,350) & \quad \mbox{if} \quad V^1 = 1\\
               N_{[0,10000]}(720,400))  & \quad \mbox{if} \quad V^1 = 0\\
               \end{array}
               \right. \\
\tilde{L}^1_0|\bar{\mathcal{D}} &\sim & N((L^1_0-671.7468)/(10\cdot352.2788)+1,0)\\
L^2_0|\bar{\mathcal{D}} &\sim &  N_{[0.06,0.8]}(0.16+0.05\cdot(L^1_0-650)/650,0.07) \\
\tilde{L}^2_0|\bar{\mathcal{D}} &\sim & N((L^2_0-0.1648594)/(10\cdot0.06980332)+1,0)\\
L^3_0|\bar{\mathcal{D}} &\sim & \left\{ \begin{array}{cl}
               N_{[-5,5]}(- 1.65 + 0.1 \cdot V^3 + 0.05 \cdot (L^1_0 - 650)/650 \\+ 0.05 \cdot (L^2_0 - 16)/16,1) & \quad \mbox{if} \quad V^1 = 1\\
               N_{[-5,5]}( -2.05 + 0.1 \cdot V^3 + 0.05 \cdot (L^1_0 - 650)/650 \\+ 0.05 \cdot (L^2_0 - 16)/16,1))  & \quad \mbox{if} \quad V^1 = 0\\
               \end{array}
               \right. \\
A_0|\bar{\mathcal{D}} &\sim& B(p=0) \\
C_0|\bar{\mathcal{D}} &\sim& B(p=0) \\
Y_0|\bar{\mathcal{D}} &\sim &  N_{[-5,5]}(-2.6 + 0.1 \cdot I(V^3 > 2) + 0.3 \cdot I(V^1 = 0) + (L^3_0 + 1.45),1.1) \\
\end{eqnarray*}

\noindent For $t>0$:

\indent\begin{eqnarray*}
L^1_t|\bar{\mathcal{D}} &\sim & \left\{ \begin{array}{cl}
               N_{[0,10000]}(13\cdot\log(t \cdot (1034-662)/8) + L^1_{t-1} + 2 \cdot L^2_{t-1}\\ + 2 \cdot L^3_{t-1} + 2.5 \cdot A_{t-1},50) & \quad \mbox{if} \quad t \in \{1,2,3,4\}\\
               N_{[0,10000]}(4\cdot\log(t \cdot (1034-662)/8) + L^1_{t-1} + 2 \cdot L^2_{t-1} \\+ 2 \cdot L^3_{t-1} + 2.5 \cdot A_{t-1},50) & \quad \mbox{if} \quad t \in \{5,6\}\\
               \end{array}
               \right. \\
L^2_t|\bar{\mathcal{D}} &\sim &  N_{[0.06,0.8]}(L^2_{t-1} + 0.0003 \cdot (L^1_t - L^1_{t-1}) + 0.0005 \cdot (L^3_{t-1}) + 0.0005 \cdot A_{t-1} \cdot \tilde{L}^1_0,0.02) \\
L^3_t|\bar{\mathcal{D}} &\sim &  N_{-5,5}(L^3_{t-1} + 0.0017 \cdot (L^1_t - L^1_{t-1}) + 0.2 \cdot (L^2_t - L^2_{t-1}) + 0.005 \cdot A_{t-1} \cdot \tilde{L}^2_0,0.5) \\
A_t|\bar{\mathcal{D}}  &\sim& \left\{ \begin{array}{cl}
               B(p=1) & \quad \mbox{if} \quad A_{t-1} = 1\\
               B(p=1/(1+\exp(-[-2.4 + 0.015 \cdot (750 - L^1_t) + 5 \cdot (0.2 - L^2_t) \\- 0.8 \cdot L^3_t + 0.8 \cdot t])))  & \quad \mbox{if} \quad A_{t-1} = 0\\
               \end{array}
               \right.  \\
C_t|\bar{\mathcal{D}} &\sim& B(p=1/(1+\exp(-[-6+ 0.01 \cdot (750 - L^1_t) + 1 \cdot (0.2 - L^2_t) - 0.65 \cdot L^3_t - A_t]))) \\
Y_t|\bar{\mathcal{D}} &\sim &  N_{[-5,5]}(Y_{t-1} +
         0.00005 \cdot (L^1_t - L^1_{t-1}) - 0.000001 \cdot \left((L^1_t  - L^1_{t-1})\cdot \sqrt{\tilde{L}^1_0}\right)^2\\
         && + 0.01 \cdot (L^2_t - L^2_{t-1})-  0.0001 \cdot \left((L^2_t - L^2_{t-1})\cdot \sqrt{\tilde{L}^2_0}\right)^2\\
         && + 0.07 \cdot ((L^3_t-L^3_{t-1})\cdot(L^3_0+1.5135)) - 0.001 \cdot ((L^3_t-L^3_{t-1})\cdot(L^3_0+1.5135))^2  \\
         && + 0.005 \cdot A_t + 0.075 \cdot A_{t-1} + 0.05 \cdot A[t] \cdot A[t-1] ,0.01) \\
\end{eqnarray*}
}}

}
\bibliographystyle{spbasic}
{\footnotesize
\bibliography{MINote}

\begin{thebibliography}{56}
\providecommand{\natexlab}[1]{#1}
\providecommand{\url}[1]{{#1}}
\providecommand{\urlprefix}{URL }
\expandafter\ifx\csname urlstyle\endcsname\relax
  \providecommand{\doi}[1]{DOI~\discretionary{}{}{}#1}\else
  \providecommand{\doi}{DOI~\discretionary{}{}{}\begingroup
  \urlstyle{rm}\Url}\fi
\providecommand{\eprint}[2][]{\url{#2}}

\bibitem[{Akaike(1973)}]{Akaike:1973}
Akaike H (1973) Information theory and an extension of the maximum likelihood
  principle. Proceeding of the Second International Symposiumon Information
  Theory Budapest pp 267--281

\bibitem[{Bang and Robins(2005)}]{Bang:2005}
Bang H, Robins JM (2005) Doubly robust estimation in missing data and causal
  inference models. Biometrics 64(2):962--972

\bibitem[{Breiman(2001)}]{Breiman:2001}
Breiman L (2001) Random forests. Machine Learning 45(1):5--32

\bibitem[{Buckland et~al(1997)Buckland, Burnham, and Augustin}]{Buckland:1997}
Buckland ST, Burnham KP, Augustin NH (1997) Model selection: an integral part
  of inference. Biometrics 53:603--618

\bibitem[{Burnham and Anderson(2002)}]{Burnham:2002}
Burnham K, Anderson D (2002) Model selection and multimodel inference. A
  practical information-theoretic approach. Springer, New York

\bibitem[{Chatfield(1995)}]{Chatfield:1995}
Chatfield C (1995) Model uncertainty, data mining and statistical inference.
  Journal of the Royal Statistical Society A 158:419--466

\bibitem[{Cheng et~al(2015)Cheng, Ing, and Yu}]{Cheng:2015}
Cheng TCF, Ing CK, Yu SH (2015) Toward optimal model averaging in regression
  models with time series errors. Journal of Econometrics 189(2):321--334

\bibitem[{Daniel et~al(2013)Daniel, Cousens, De~Stavola, Kenward, and
  Sterne}]{Daniel:2013}
Daniel RM, Cousens SN, De~Stavola BL, Kenward MG, Sterne JA (2013) Methods for
  dealing with time-dependent confounding. Statistics in Medicine
  32(9):1584--618

\bibitem[{Draper(1995)}]{Draper:1995}
Draper D (1995) Assessment and propagation of model uncertainty. Journal of the
  Royal Statistical Society B 57:45--97

\bibitem[{Fletcher and Dillingham(2011)}]{Fletcher:2011}
Fletcher D, Dillingham PW (2011) Model-averaged confidence intervals for
  factorial experiments. Computational Statistics and Data Analysis
  55:3041--3048

\bibitem[{Friedman et~al(2010)Friedman, Hastie, and Tibshirani}]{Friedman:2010}
Friedman J, Hastie T, Tibshirani R (2010) Regularization paths for generalized
  linear models via coordinate descent. Journal of Statistical Software
  33(1):1--22

\bibitem[{Gao et~al(2016)Gao, Zhang, Wang, and Zou}]{Gao:2016}
Gao Y, Zhang XY, Wang SY, Zou GH (2016) Model averaging based on
  leave-subject-out cross-validation. Journal of Econometrics 192(1):139--151

\bibitem[{Gelman and Su(2016)}]{Gelman:2016}
Gelman A, Su YS (2016) arm: Data Analysis Using Regression and
  Multilevel/Hierarchical Models.
  \urlprefix\url{https://CRAN.R-project.org/package=arm}, {R} package version
  1.9-3

\bibitem[{Gruber and van~der Laan(2012)}]{Gruber:2012}
Gruber S, van~der Laan MJ (2012) tmle: An {R} package for targeted maximum
  likelihood estimation. Journal of Statistical Software 51(13):1--35

\bibitem[{Hansen(2007)}]{Hansen:2007}
Hansen BE (2007) Least squares model averaging. Econometrica 75:1175--1189

\bibitem[{Hansen(2008)}]{Hansen:2008}
Hansen BE (2008) Least squares forecast averaging. Journal of Econometrics
  146:342--350

\bibitem[{Hansen and Racine(2012)}]{Hansen:2012}
Hansen BE, Racine J (2012) Jackknife model averaging. Journal of Econometrics
  167:38--46

\bibitem[{Hjort and Claeskens(2003)}]{Hjort:2003}
Hjort L, Claeskens G (2003) Frequentist model average estimators. Journal of
  the American Statistical Association 98:879--945

\bibitem[{Hoeting et~al(1999)Hoeting, Madigan, Raftery, and
  Volinsky}]{Hoeting:1999}
Hoeting JA, Madigan D, Raftery AE, Volinsky CT (1999) Bayesian model averaging:
  a tutorial. Statistical Science 14:382--417

\bibitem[{Kabaila et~al(2016)Kabaila, Welsh, and Abeysekera}]{Kabaila:2016}
Kabaila P, Welsh A, Abeysekera W (2016) Model-averaged confidence intervals.
  Scandinavian Journal of Statistics 43:35--48

\bibitem[{Van~der Laan and Petersen(2007)}]{vanderLaan:2007}
Van~der Laan M, Petersen M (2007) Statistical learning of origin-specific
  statistically optimal individualized treatment rules. International Journal
  of Biostatistics 3:Article 3

\bibitem[{Van~der Laan and Rose(2011)}]{vanderLaan:2011}
Van~der Laan M, Rose S (2011) Targeted Learning. Springer

\bibitem[{Van~der Laan et~al(2008)Van~der Laan, Polley, and
  Hubbard}]{vanderLaan:2008}
Van~der Laan M, Polley E, Hubbard A (2008) Super learner. Statistical
  Applications in Genetics and Molecular Biology 6:Article 25

\bibitem[{Leeb and P{\"o}tscher(2005)}]{Leeb:2005}
Leeb H, P{\"o}tscher BM (2005) Model selection and inference: facts and
  fiction. Econometric Theory 21:21--59

\bibitem[{Leeb and P{\"o}tscher(2008)}]{Leeb:2008}
Leeb H, P{\"o}tscher BM (2008) Model Selection, Springer, New York, pp 785--821

\bibitem[{Lendle et~al(2017)Lendle, Schwab, Petersen, and {van der
  Laan}}]{Lendle:2017}
Lendle SD, Schwab J, Petersen ML, {van der Laan} MJ (2017) {ltmle}: An {R}
  package implementing targeted minimum loss-based estimation for longitudinal
  data. Journal of Statistical Software 81(1):1--21

\bibitem[{Liang et~al(2011)Liang, Zou, Wan, and Zhang}]{Liang:2011}
Liang H, Zou GH, Wan ATK, Zhang XY (2011) Optimal weight choice for frequentist
  model average estimators. Journal of the American Statistical Association
  106(495):1053--1066

\bibitem[{Liu and Kuo(2016)}]{Liu:2016}
Liu C, Kuo B (2016) Model averaging in predictive regressions. Econometrics
  Journal 19(2):203--231

\bibitem[{Liu et~al(2016)Liu, Okui, and Yoshimura}]{Liu:2016b}
Liu QF, Okui R, Yoshimura A (2016) Generalized least squares model averaging.
  Econometric Reviews 35(8-10):1692--1752

\bibitem[{Mallows(1973)}]{Mallows:1973}
Mallows C (1973) Some comments on {$C_p$}. Technometrics 15:661--675

\bibitem[{Petersen et~al(2014)Petersen, Schwab, Gruber, Blaser, Schomaker, and
  van~der Laan}]{Petersen:2014}
Petersen M, Schwab J, Gruber S, Blaser N, Schomaker M, van~der Laan M (2014)
  Targeted maximum likelihood estimation for dynamic and static longitudinal
  marginal structural working models. Journal of Causal Inference 2:147--185

\bibitem[{Polley et~al(2017)Polley, LeDell, Kennedy, and {van der
  Laan}}]{Polley:2017}
Polley E, LeDell E, Kennedy C, {van der Laan} M (2017) SuperLearner: Super
  Learner Prediction.
  \urlprefix\url{https://CRAN.R-project.org/package=SuperLearner}, {R} package
  version 2.0-22

\bibitem[{P{\"o}tscher(2006)}]{Poetscher:2006}
P{\"o}tscher B (2006) The distribution of model averaging estimators and an
  impossibility result regarding its estimation. In: Ho H, Ing C, Lai T (eds)
  IMS Lecture Notes: Time series and related topics, vol~52, pp 113--129

\bibitem[{Raftery et~al(2017)Raftery, Hoeting, Volinsky, Painter, and
  Yeung}]{Raftery:2017}
Raftery A, Hoeting J, Volinsky C, Painter I, Yeung KY (2017) {BMA}: Bayesian
  Model Averaging. \urlprefix\url{https://CRAN.R-project.org/package=BMA}, {R}
  package version 3.18.7

\bibitem[{Rao et~al(2001)Rao, Wu, Konishi, and Mukerjee}]{Rao:2001}
Rao CR, Wu Y, Konishi S, Mukerjee R (2001) On model selection. Lecture
  Notes-Monograph Series 38:1--64

\bibitem[{Robins and Hernan(2009)}]{Robins:2009}
Robins J, Hernan MA (2009) Estimation of the causal effects of time-varying
  exposures. In: Fitzmaurice G, Davidian M, Verbeke G, Molenberghs G (eds)
  Longitudinal Data Analysis, CRC Press, pp 553--599

\bibitem[{Sala-I-Martin et~al(2004)Sala-I-Martin, Doppelhofer, and
  Miller}]{salaimartin:2004}
Sala-I-Martin X, Doppelhofer G, Miller RI (2004) Determinants of long-term
  growth: A {B}ayesian averaging of classical estimates (bace) approach.
  American Economic Review 94(4):813--835

\bibitem[{Schomaker(2012)}]{Schomaker:2012}
Schomaker M (2012) Shrinkage averaging estimation. Statistical Papers
  53(4):1015--1034

\bibitem[{Schomaker(2017{\natexlab{a}})}]{Schomaker:2017}
Schomaker M (2017{\natexlab{a}}) {MAMI}: Model Averaging (and Model Selection)
  after Multiple Imputation. R package version 0.9.10

\bibitem[{Schomaker(2017{\natexlab{b}})}]{Schomaker:2017b}
Schomaker M (2017{\natexlab{b}}) Model Averaging and Model Selection after
  Multiple Imputation using the {R}-package MAMI.
  \urlprefix\url{http://mami.r-forge.r-project.org}

\bibitem[{Schomaker and Heumann(2014)}]{Schomaker:2014}
Schomaker M, Heumann C (2014) Model selection and model averaging after
  multiple imputation. Computational Statistics and Data Analysis 71:758--770

\bibitem[{Schomaker and Heumann(2018)}]{Schomaker:2018}
Schomaker M, Heumann C (2018) Bootstrap inference when using multiple
  imputation. Statistics in Medicine 37(14):2252--2266

\bibitem[{Schomaker et~al(2016)Schomaker, Davies, Malateste, Renner, Sawry,
  N'Gbeche, Technau, Eboua, Tanser, Sygnate-Sy, Phiri, Amorissani-Folquet, Cox,
  Koueta, Chimbete, Lawson-Evi, Giddy, Amani-Bosse, Wood, Egger, and
  Leroy}]{Schomaker:2016}
Schomaker M, Davies MA, Malateste K, Renner L, Sawry S, N'Gbeche S, Technau K,
  Eboua FT, Tanser F, Sygnate-Sy H, Phiri S, Amorissani-Folquet M, Cox V,
  Koueta F, Chimbete C, Lawson-Evi A, Giddy J, Amani-Bosse C, Wood R, Egger M,
  Leroy V (2016) Growth and mortality outcomes for different antiretroviral
  therapy initiation criteria in children aged 1-5 years: A causal modelling
  analysis from {W}est and {S}outhern {A}frica. Epidemiology 27:237--246

\bibitem[{Sofrygin et~al(2017)Sofrygin, {van der Laan}, and
  Neugebauer}]{Sofrygin:2016}
Sofrygin O, {van der Laan} MJ, Neugebauer R (2017) {simcausal} {R} package:
  Conducting transparent and reproducible simulation studies of causal effect
  estimation with complex longitudinal data. Journal of Statistical Software
  81(2):1--47

\bibitem[{Tibsharani(1996)}]{Tibshirani:1996}
Tibsharani R (1996) Regression shrinkage and selection via the lasso. Journal
  of the Royal Statistical Society B 58:267--288

\bibitem[{Turek and Fletcher(2012)}]{Turek:2012}
Turek D, Fletcher D (2012) Model-averaged wald confidence intervals.
  Computational Statistics and Data Analysis 56:2809--2815

\bibitem[{Wan et~al(2010)Wan, Zhang, and Zou}]{Wan:2010}
Wan ATK, Zhang X, Zou GH (2010) Least squares model averaging by {M}allows
  criterion. Journal of Econometrics 156:277--283

\bibitem[{Wang and Zhou(2012)}]{Wang:2012}
Wang H, Zhou S (2012) Interval estimation by frequentist model averaging.
  Communications in Statistics -- Theory and Methods 42(23):4342--4356

\bibitem[{Wang et~al(2009)Wang, Zhang, and Zou}]{Wang:2009}
Wang H, Zhang X, Zou G (2009) Frequentist model averaging: a review. Journal of
  Systems Science and Complexity 22:732--748

\bibitem[{Wood(2006)}]{Wood:2006}
Wood SN (2006) Generalized additive models: an introduction with R. Chapman and
  Hall/CRC

\bibitem[{Yan(2007)}]{Yan:2007}
Yan J (2007) Enjoy the joy of copulas: with package copula. Journal of
  Statistical Software 21:1--21

\bibitem[{Zhang and Liu(2017)}]{Zhang:2017}
Zhang X, Liu CA (2017) {Inference after Model Averaging in Linear Regression
  Models}. IEAS Working Paper : academic research 17-A005, Institute of
  Economics, Academia Sinica, Taipei, Taiwan,
  \urlprefix\url{https://ideas.repec.org/p/sin/wpaper/17-a005.html}

\bibitem[{Zhang et~al(2014)Zhang, Zou, and Liang}]{Zhang:2014}
Zhang XY, Zou GH, Liang H (2014) Model averaging and weight choice in linear
  mixed-effects models. Biometrika 101(1):205--218

\bibitem[{Zhang et~al(2015)Zhang, Zou, and Carroll}]{Zhang:2015}
Zhang XY, Zou GH, Carroll RJ (2015) Model averaging based on kullback-leibler
  distance. Statistica Sinica 25(4):1583--1598

\bibitem[{Zhang et~al(2016{\natexlab{a}})Zhang, Ullah, and Zhao}]{Zhang:2016b}
Zhang XY, Ullah A, Zhao SW (2016{\natexlab{a}}) On the dominance of mallows
  model averaging estimator over ordinary least squares estimator. Economics
  Letters 142:69--73

\bibitem[{Zhang et~al(2016{\natexlab{b}})Zhang, Yu, Zou, and
  Liang}]{Zhang:2016}
Zhang XY, Yu DL, Zou GH, Liang H (2016{\natexlab{b}}) Optimal model averaging
  estimation for generalized linear models and generalized linear mixed-effects
  models. Journal of the American Statistical Association 111(516):1775--1790

\end{thebibliography}
}

\end{document}